\documentclass[aps,prd,preprintnumbers,superscriptaddress,showpacs,nofootinbib,12pt,floatfix]{revtex4}
\usepackage{graphicx}
\usepackage{dcolumn}
\usepackage{bm}
\usepackage{amsmath}
\usepackage{amssymb}
\def\[{\left[}
\def\]{\right]}

\def\gsim{\lower.7ex\hbox{$\;\stackrel{\textstyle>}{\sim}\;$}}
\def\lsim{\lower.7ex\hbox{$\;\stackrel{\textstyle<}{\sim}\;$}}
\def\sq2{{\sqrt{2}}}

\def\DRED{\ifmmode{{\rm DRED}} \else{{DRED}} \fi}

\newcommand{\beq}{\begin{equation}}
\newcommand{\eeq}{\end{equation}}
\newcommand{\beqa}{\begin{eqnarray}}
\newcommand{\eeqa}{\end{eqnarray}}

\begin{document}
\begin{flushright}
DUAL-CP-07-05\\
\today
\end{flushright}
\title{Holographic Dark Matter and Higgs }
\author{J. Lorenzo D\'{\i}az-Cruz}
\email{jldiaz@fcfm.buap.mx} \affiliation{Fac. de Cs.
F\'{\i}sico-Matem\'aticas, BUAP,\\ Apdo. Postal 1364,
Puebla, Pue., C.P. 72000, M\'exico,  \\ 
and Dual CP Institute of High Energy Physics}

\date{\today}
\begin{abstract}
We identify possible dark matter candidates within the class of strongly
interacting models where electroweak symmetry breaking is triggered by
a light composite Higgs boson.  In these models, the Higgs 
boson emerges as a Holographic pseudo-goldstone boson, while dark
matter can be identified as a fermionic composite state $X^0$,
which is made stable through a conserved (``dark'') quantum number. 
An effective lagrangian description of both the Higgs and dark 
matter is proposed, that includes higher-dimensional operators 
suppressed by an scale $\Lambda_i$. These operators will induce 
deviations from the standard Higgs properties that could be 
meassured at future colliders (LHC,ILC), and thus provide information on
the dark matter scale.  The dark matter $X^0$, is expected to
have a mass of order $m_{X^0} \lsim 4\pi f\simeq O(TeV)$, which is
in agreement with the values extracted from the cosmological bounds 
and the experimental searches for dark matter.
\end{abstract}
\pacs{12.60.Cn,12.60.Fr,11.30.Er}
\maketitle

\newpage

{\bf{1.- Introduction.}} Explaining the nature of electroweak symmetry
breaking (EWSB) and dark matter (DM) have become two of the most important
problems in modern elementary particle physics and cosmology today
\cite{EWSBrev, COSMOrev}. Within the standard model (SM), electroweak precision
tests prefer the existence of a light Higgs boson, with a mass of order of 
the electroweak (EW) scale $v\simeq 175$ GeV, that
should be tested soon at the LHC \cite{hixlhc}. Similarly, plenty of
astrophysical and cosmological data points towards the 
existence of a dark matter component, that accounts for about 12\% 
of the matter-energy content of our universe \cite{DMrev}. 
A weakly-interacting massive particle (WIMP) with a mass of the order
of the EW scale, seems a most viable option for the dark matter.
 What is the nature of EWSB and dark matter, and how do they fit in
our current understanding of elementary particles, is
however not  known.

Given the similar requirements on masses and interactions for both
particles, Higgs and DM, one can naturally ask whether they could share a
common origin. Within the minimal SUSY SM (MSSM) \cite{mssmrev}, 
which has become one of the most popular extensions of the SM, 
there are several WIMP candidates (neutralino, sneutrino, gravitino) \cite{mssmXDM}.
Among them, the neutralino has been most widely studied; it
is a combination of the so called Higgsinos and gauginos, which are the
SUSY partners of the Higgs and gauge bosons. Thus, in the SUSY case, 
the fermion-boson symmetry provides the connection between EWSB 
and Dark Matter.
However, many new  models have been proposed more recently \cite{ewsbrev},
which provide alternative theoretical foundation to stabilize the Higgs
mechanism. Some of these new models assume that EWSB originates in a
new strongly interacting sector, and have been originally motivated by 
the studies of extra dimensions \cite{exdimrev}. In some of them, 
candidates for dark matter have been proposed, such as the lightest 
T-odd particle (LTP) within little Higgs models \cite{LTP}  or 
the lightest KK particle (LKP) in models with universal 
extra-dimensions \cite{LKP}.

In this letter, we are interested in searching for possible dark
matter candidates, within the so called Holographic Higgs  models
\cite{HoloHiggs}. Here, EWSB is triggered by a light composite 
Higgs boson, which emerges as a pseudo-goldstone boson. 
Within this class of  strongly interacting models,
we shall propose that a stable  composite ``Baryon'', can account 
for the dark matter.  This particle can be made stable by impossing 
a conserved (Dark) quantum  number, and it will be denoted as the 
lightest Holographic fermionic particle ($LHP$). 
The effective lagrangian description of both the Higgs and dark
matter, includes higher-dimensional operators suppressed by an scale
$\Lambda_i$ ($i=H,X$), which will induce deviations from the SM
predictions for the Higgs properties. Thus, meassuring these effects at 
future colliders (LHC,ILC), could also provide information on
the dark matter scale. Furthermore, 
it is likely that because of compositeness, the Higgs boson will be heavier 
than in the SM case, as it can be derived from EW precision tests. 
 Having SM interactions, the LHP will share similar characteristics 
with other WIMP candidates,  however the composite nature of $X^0$
will also have important  implications for cosmological bounds and the 
experimental searches for dark matter.

 This picture, where strong interactions produce a light pseudo-goldstone
boson and a heavier stable fermion, is not strange at all in nature.
 This is precisely what happens in ordinary hadron physics, where
the pion and the proton play such roles, namely they appear as two-
and three-quark bound states, formed by the action of the strong  
QCD interaction. In this paper we shall propose models that produce
a similar pattern for the Higgs and dark matter, but at a higher 
energy scale, and with a stable neutral state instead of a charged one. 
We believe that such scenario is very attractive and unifying,
and it could provide further understanding of both EWSB and DM problems.
Although we shall formulate our ideas using a generic 
effective lagrangian approach, we shall also discuss specific models 
within the known  Holographic Higgs models \cite{HoloHiggs}.

{\bf{2.- Holographic Higgs and Dark Matter.}} We are thus interested
in looking for a dark matter candidate, within the context of strongly 
interacting models that produce  a light composite Higgs boson. Although these
models admit a dual AdS/CFT description, we shall discuss its main 
features from the 4D point of view, ocacionally relaying on the
corresponding 5D description to clarify some issues.
 From the 4D perspective, these models are formulated 
through an effective lagrangian  \cite{EffholoHIX,EffholoKK},
that includes two sectors: 
i) The SM sector that contains the gauge bosons and 
most of the quarks and leptons, which is characterized by a 
generic coupling $g_{sm}$ (gauge or Yukawa), 
and ii) A new strongly interacting sector, characterized
by another  coupling  $g_{*}$ and an scale $M_R$. This scale can be associated
with the mass of the lowest composite resonance, which corresponds
to the lightes KK resonance in the dual 5D Ads description; in
ordinary QCD it is the mass of the rho meson ($\rho$). The couplings
are choosen to satisfy  $g_{sm} \lsim g_{*} \lsim 4 \pi$. 
As a result of the dynamics and global symmetries of the strongly
interacting sector, a composite Higgs boson emerges as an  exactly 
massless goldstone boson in the limit $g_{sm} \to 0$. 
SM interactions then produce a deformation of the theory, and the
Higgs boson becomes massive.  Thus, radiative effects induce a 
Higgs mass, which can be written as:
$m_h \simeq (\frac{g_{sm}}{4\pi}) M_R$.

Like the Higgs, we propose that dark matter arises as a composite 
states from the strongly interacting sector; in fact, a whole tower of
fermionic states $X^0, X^{\pm}, X{ \pm,\pm}...$ should appear.
Similarly to what happens in ordinary QCD, where the proton is 
stable because of Baryon number conservation, we shall also assume 
that the lightest Holographic fermionic particle (LHP) $X^0$ will
be stable because a new conserved quantum number, that we call 
``Dark Number'' ($D_N$). 
Thus, the SM particles and the ``Mesonic'' states, like 
the Higgs boson,  will have zero Dark number ($D_N(SM)=0$), 
while the ``baryonic'' states like $X^0$,  will have +1
dark number ($D_N(X^0)=+1$).
 For a deformed  $\sigma$ type model of the strongly interacting 
sector, one can use  NDA to derive a bound 
on the mass of $X^0$, namely  $m_{X^0} \lsim 4 \pi f$, where $f$ is
the analogue of the pion decay constant.  
 It is usualy assumed that lightest resonance of the Holographic 
theory, corresponds to  a vectorial resonance, in analogy with
ordinary QCD. However, because we lack a detailed quantitative 
understanding of the strongly interacting theory, we admit the
possibility that $X^0$ corresponds to the lightest state. 
Thus, the natural value for $M_{X^0}$ will be in the TeV range, 
somehow heavier than the SUSY candidates for dark matter.

The properties of the Holographic Higgs boson can be described in
terms of the following effective lagrangian:
\begin{equation}
{\cal{L}}_H = {\cal{L}}^H_{sm} + \sum \frac{\alpha_i}{(\Lambda_i)^{n-4}} O_{in}
\end{equation} 
where ${\cal{L}}^H_{sm}$ denotes the SM Higgs lagrangian.
 The next term contains higher-dimensional operators $O_{in}$ ($n\geq 6$),
that can induce deviations from the SM for the Higgs properties. 
The coefficient $\alpha_i$ will depend on gauge/Yukawa couplings, 
mixing angles and possible loop factor, while the scale
$\Lambda_i$ could be either $f$ or $M_R$, depending on the nature of
each operator. Examples of such operators include: 
$O_W=i(H^\dagger \sigma^i D^\mu H)(D^\nu W_{\mu\nu})^i$,
$O_B=i(H^\dagger \sigma^i D^\mu H)(\partial^\nu B_{\mu\nu})$,
$O_{HW}=i(D^\mu H)^\dagger \sigma^i (D^\nu H) W^i_{\mu\nu}$,
$O_{HB}=i(D^\mu H)^\dagger (D^\nu H) B_{\mu\nu}$,
$O_{T}=i(H^\dagger D^\mu H)(H^\dagger D_\mu H)$,
$O_H =i \partial^\mu (H^\dagger H) \partial_\mu (H^\dagger H)$, 
 as discussed in ref. \cite{EffholoHIX}.
These operators have been studied in the past  for the most  
general effective lagrangian extension of the SM, and they 
can induce, for instance, modifications to the  SM bounds 
on the Higgs mass obtained from EW precision tests 
(EWPT) \cite{ourHMetal}. In particular,
the operator $O_T$ can help to increasse the limit on the Higgs mass 
above 300 GeV, for relatively natural values of 
parameters, i.e. with $\alpha_i=O(1)$ and $\Lambda_H \simeq 1$ TeV. 
Furthermore, at LHC it will be possible to meassure the corrections 
to the Higgs couplings induced by those operators, with a precision that will 
translate into bounds on the scale $M_R$ of the order 5-7 TeV, while
at ILC this range will extend up to about 30 TeV \cite{EffholoHIX}. 
It is important to stress that such analysis would be re-interpreted as an
alternative method to derive indirect constraints on the DM scale.

{\bf{3.- Holographic Dark Matter models.}}
Our proposed dark matter candidate ($LHP$) can arise in any of the 
Holographic Higgs models proposed so far; we shall argue that it is a generic
feature of this class of strongly interacting theories \cite{HoloHiggs}. 
However, its specific realization will depend on the particular model under
consideration, which will fix the quantum numbers of the LHP.
In this paper, we shall consider the simplest possibilities for the
LHP,  within the models discussed in \cite{HoloHiggs,holohixM}. 
From the 4D perspective, each model is defined by impossing
a global symmetry $G$ on the strongly interacting 
sector, of which only the SM subgroup $H=SU(2) \times U(1)$ will be gauged. 
Thus, the DM model will be difined by specifying a $G$-multiplet,
which is composed of an $H$-multiplet that has SM gauge interactions,
plus some extra singlets. We call {\it{Active DM}} those cases when
the LHP belongs to the $H$-multiplet, while {\it{Sterile  DM}} will be
used for models where the LHP is a SM singlet. 

Let us consider first the models that can be constructed  with 
$G=SU(3) \times U(1)_X$. The extra $U(1)_X$  is needed in order to get 
the correct assignment for SM hypercharges. Under $SU(3) \times U(1)_X$
the SM doublets and d-type singlets are included in $SU(3)$ triplets, 
namely: the SM quark doublet appears in:  $Q \simeq 3^*_{1/3}$, while the
d-type singlet is contained in: $D \simeq 3_0$. The SM up-type singlet is
contained in a TeV-brane field that transforms as a singlet: 
$U \simeq 1_{1/3}$. The SM hypercharge for the fermions is
obtained from the relation: $Y=\frac{T_8}{\sqrt{3}}+X$, while
the electric charge arises from the usual relation:  
$Q_{em}=T_3+ \frac{Y}{2}$.
The additional composite fermions (``Baryons''), which shall contain
the  LHP, must also appear in complete multiplets of $SU(3)$,
in order to keep under control their radiative contributions 
to the  Higgs mass \cite{lightCUST}. Furthermore, 
admiting only the lowest dimensional representations under $SU(3)$ 
(triplets and singlets)  to accomodate an electrically neutral LHP 
candidate, requires: $X= \pm 1/3, \pm 2/3$, which admit
SM singlets and doublets. 
A classification of the corresponding  active and sterile Holographic 
dark matter models is listed in Table 1. It includes, for instance, the case of
an $SU(3)$ anti-triplet with $X=1/3$, which can be written as: 
$\Psi= (N^0_1, C^+_1,N^0_2)^T$.
Therefore, we can have two options for the LHP: 
i) Model 1 (active)
where the LHP is part of the SM doublet $\psi=  (N^0_1, C^+_1)$, 
i.e. $X^0=N^0_1$, similar to a heavy neutrino, and 
ii) Model 2 (sterile) where the LHP is a SM singlet, i.e. $X^0=N^0_2$.
In this case $X^0$ does not have SM couplings at tree-level, but they 
could appear through the inclusion of higher-dimensional operators. 
 Allowing the inclusions of SU(3) octets leads to the possibility of
having also SM triplets with $Y=0$ (model 7). On the other hand, 
when one considers the Higgs model with $G=SO(5) \times U(1)_X$
\cite{lightCUST}, there is number of other posibilities 
for the quantum numbers of the $G-$ and $H-$multiplets.
In particular, one could have a DM neutral state belonging to
SM Doublet, SM triplet, SM singlet, etc.
A detailed study of these DM option will be carried in a forthcoming 
publication \cite{ournextDM}. Here we shall only determine the
viability of the $SU(3)$ models listed in table 1.

 The couplings of $X^0$ with the SM sector will include
both renormalizable and effective interactions. The renormalizable
interactions will be fixed by the quantum numbers of $X^0$, 
while the effective lagrangian will include higher-dimensional 
operators, which would represent both the effects from the 
integration of heavy fermions that belong to the $G-$multiplet, 
as well as the composite nature of LHP. Thus, we write the
full lagrangian for DM as follows:
\begin{equation}
{\cal{L}}_{DM} = \bar{X}^0 (\gamma^\mu D_\mu - M_x) X^0
+ \sum \frac{\alpha_i}{(\Lambda_i)^{n-4}} O_{in}
\end{equation} 
where $D_\mu= \partial_\mu -i g_{x} T^i W^i_\mu- g'_{x} \frac{Y}{2} B_\mu$.
For the case with $Y=0$, we have $g_x(g'_{x})=g_2 (g_1)$, while for $Y=1$ it
can be a different story, as it will be discussed next. 

\begin{table}[t!] 
\begin{center}  
\begin{tabular}{| c| c| c| c| }  
\hline\hline  
$U(1)_X$  & $G$-multiplets & $H$-multiplets & LHP models \\  
\hline  
$+\frac{1}{3}$  & $\bf{3}^*$:  $\Psi_1=(N^0_1,C^+_1,N^0_2)^T$ &   
 $\bf{2}^*$: $\psi_1=(N^0_1,C^+_1)^T$ & 1) $X^0=N^0_1$ (Active) \\
              &     &       & 2) $X^0=N^0_2$ (Sterile)  \\
\hline  
$-\frac{1}{3}$  & $\bf{3}^*$: $\Psi_2=(C^-_2,N^0_3,N^0_4)^T$ &   
  $\bf{2}^*$: $\psi_2=(C^-_2,N^0_3)^T$ & 3) $X^0=N^0_3$ (Active) \\
              &     &       & 4) $X^0=N^0_4$ (Sterile)  \\
\hline 
$+\frac{2}{3}$  & $\bf{3}^*$: $\Psi_3=(N^0_5,C^+_3,C^+_4)^T$ &   
  $\bf{2}^*$: $\psi_3=(N^0_5,C^+_3)^T$ & 5) $X^0=N^0_5$ (Active)\\
\hline 
$-\frac{2}{3}$  & $\bf{3}^*$: $\Psi_4=(C^-_5,N^0_6,C^-_6)^T$ &   
 $\bf{2}^*$: $\psi_4=(C^-_5,N^0_6)^T$ & 6) $X^0=N^0_6$ (Active) \\
\hline 
$+x$  & $\bf{8}$ \, : $\Psi_5=$full octet mult. &   
 $\bf{3}$: $\psi_5=(C^+_7, N^0_7,C^-_7)^T$ & 7) $X^0=N^0_7$ (Active) \\
\hline 
    
\end{tabular}  
\end{center}  
\vspace{-0.55cm} 
\caption{ 
LHP candidates within the $SU(3) \times U(1)_X$ Holographic Higgs
models}  
\vspace{-0.4cm} 
\end{table} 

{\bf{4.- Holographic Dark Matter constraints.}}
We would like to discuss possible effects from the dark 
matter LHP, including its composite nature. Namely, we are interested in 
studying how to constrain the effective lagrangian (2),
using both cosmology and the experimental searches for DM.
Three cases of models shown in table 1 will be analyzed here. Namely:
i) Active LHP models with $Y\neq 0 $, ii) Active LHP models with $Y=0$,
and iii) Sterile LHP models. 
We shall discuss first the calculation for the relic density of 
DM. After including the interaction with SM gauge bosons, 
the result for the relic density calculation can be written
in terms of the thermal averaged cross-section $<\sigma v>$ as: 
\begin{equation} 
\Omega_X h^2 = \frac{2.57\times 10^{-10} }{ <\sigma v>}=
\frac{ 2.57\times 10^{-10} M^2_X}{C_{T,Y} }  
\end{equation} 
where  the constant $C_{T,Y}$ depends on the isospin (T) and
hypercharge (Y) of the LHP candidate. Numerical values for
$C_{T,Y}$ for the lowes-dimensional representations are:  
$C_{1/2,1/2}=0.004 $,  $C_{1,0}= 0.01$
In order to have agreement with current data on relic DM 
density, i.e. $\Omega_X h^2 =0.11\pm 0.066$ \cite{wmapA},
model 1 requires $M_X=1.3$ TeV, while model 7
requires $M_X=2.1$ TeV. It is quite remarkable that these are
precisely the right mass values expected in the strongly interacting
theory!. 

Constraints on the LHP candidates can also be derived from the 
interpretation of experimental searches for dark matter signals. 
We shall discuss here the direct search for DM based on the 
nucleon-LHP scattering \cite{CDMexp}. Again, we can calculate the 
cross-section for this reaction, taking into account the LHP 
interactions with SM fields. We find that the cross section can be 
expressed as: $ \sigma_{T,Y} = \frac{G^2_F}{2\pi} f_N Y^2$, 
where $f_N$ is a factor that depends on the type of nucleus used 
in the reaction. As it was discussed in ref. \cite{MinDM}, 
vector-like dark matter with $Y=1$ is severely constrained by the
direct search, unless its coupling with the Z boson 
is suppressed with respect to the SM strength.
A suppression of this type can be realized in a natural manner for 
Holographic dark matter models. For this, we follow ref. \cite{EffholoKK}, 
and admit a possible mixing between the composite LHP
and some elementary fields having the same SM quantum numbers.
Then, the vertex $ZXX$ will be suppressed by the mixing angles
needed to go from the weak eigenstates to the physical mass eigenstates. 
To discuss an specific model, we shall consider model 1 from
Table 1, i.e. the active DM appears in a doublet 
$\psi_1=  (N^0_1, C^+_1)$. Including the elementary copy of these
fields, allows to suppress the vertex $ZXX$, which can be written as:
$\Gamma_{ZXX}= \frac{\eta g_2}{2c_W} \gamma^\mu$, 
with $\eta <1$ taking into account the mixing betwen the
elementary and composite sectors. In this case the cross-section for
$DM+N \to DM+N$ can be written as:
$\sigma = \frac{G^2_F}{2\pi} f_N {\eta}^2$.
Then, agreement with current bounds \cite{CDMexp} requires
to have $|\eta| \leq 10^{-4}$, which in turn translates into
a bound $\Lambda_i > 10$ TeV.

Finally, we discuss the sterile LHP case, considering the 
model 2. In this case, the couplings of the LHP with SM fields, 
only appear through higher-dimensional operators. 
The whole tower of dim-6 operators, includes, for instance, 
the following operator:
\begin{equation} 
O_6 = \frac{ic_x}{f^2} (H^\dagger D_\mu H) {\bar{X}} \gamma^\mu X  
\end{equation} 
where $D_\mu$ denotes the SM covariant derivative, with $c_x$
parametrizing the strength of this contribution;
this operator will induce the vertex $Z X^0 X^0$.
 Inclusing the effect from those operators that do not modify
the Lorentz vectorial structure of the vertex $Z X^0 X^0$, 
allows us to  write it as: 
$\Gamma_{ZXX}= \frac{g}{2c_W} \eta' \gamma^\mu$, 
with $\eta'$ being a parameter that measures the strength
of those new effects associated with the whole tower of such
operators; if only the operator (3) is included we have:
$\eta'=2 c_x g c_w v^2/f^2$.
Then, requiring 
$\Omega_X h^2 \simeq \Omega_{DM} h^2 = 0.11\pm 0.006$ \cite{wmapA}, 
one obtains a constraint of order: $\eta'\simeq O(0.1)$.
On the other hand, we find that the corresponding cross section for
nucleon-LHP scattering is  suppressed enough to satisfy the
experimental limits \cite{CDMexp}.

 Other experimental searches can be discussed similarly, such as the
anhihilation into photon pairs, i.e. $XX \to \gamma \gamma$.
Even more exotic signatures of this model, can be obtained by considering the
extra particles appearing in the G-multiplets, i.e. we
can look for effects form the G- or H-partners of the LHP. Within the
Holographic $SU(3)$ model 1, the LHP appears in a weak doublet, with
an extra charged state $X^-$. Because of EWPT, in particular their
contribution to the $\rho$ parameter, the mases of both particles
$X^0$ and $X^-$ should not differ by much. Thus, it should be possible to
produce pairs  $X^-X^+$ at the LHC, which will decay predominantly into 
$X^\pm \to W^\pm + X^0$. Furthermore, in a strongly interacting theory
there should be resonances of these states, which could be searched at
LHC too.
Turning now to Astrophysical signals, we could imagine that $X^-$ and
other resonances, could be produced at places with high concentrations
of dark matter, where we would observe high-energy activity.  Good 
candidates for such places, are the AGN. The high-energy
signals arising from the decays of $X^-$ into $X^0+W-$, would lead to
the prediction of cosmic rays with energies in the multi-TeV range. 
 An extensive discussion of these searches, including
the whole tower of dim-6 operators, will be presented in an extended
version of this letter \cite{ournextDM}.

{\bf{5.- Conclusions.}}
We have proposed new dark matter candidates,  within the context of 
strongly interacting Holographic Higgs models. These LHP candidates
are identified as composite fermionic states ($X^0$),  
with a mass of order $m_{X^0} \lsim 4\pi f$,  which is made
stable by assuming the existence of a conserved ``dark'' quantum
number. Thus, we suggest that there exists a connection 
between two of the most important problems in particles physics and 
cosmology: EWSB and DM.
 In these models, the Higgs boson couplings receive potentially 
large corrections, which could be tested at the coming (LHC) and 
future colliders (ILC). Measuring these deviations from SM 
predictions, will not only constrain the Higgs properties,  
but it could also provide information on the dark matter scale. 
In particular, LHC could provide indirect evidence of dark matter for 
masses of order 5-7 TeV, while ILC will be able to reach masses of
order 30 TeV. A correlated dark matter signal with
these masses should be also observed at LHC.
A list of some of the models that can appear within the $SU(3)$ 
Holographic Higgs model are shown in table 1.

 We have verified that the calculation of the LHP relic abundance, 
including the corrections to its couplings, satisfies the astrophysical
observations.  Furthermore, the current bounds on dark matter
experimental searches, such as those based on LHP-nucleon scattering,
provides stringentconstraints on the parameters of the model.
We  conclude that most favorable models
are the sterile ones with $Y=0$, like model 7 of table 1.
Although models with $Y\neq 0 $ are less favored, we 
identified a possible mechanisms within the Holographic approach,
which can help to improve their consistency.
Additional astrophysical signals from these models, can be discused 
too; for instance, one can look for the production of excited states 
or the G- or H-partners of $X^-$, which ccould produce
TeV-scale High energy cosmic rays that could be searched in 
future experiments\cite{miniHAWK}.


Acknowledgements: Discussions with A. Aranda, H. Salazar and the
Dual-CP Dorados are sincerely acknowledged.
This work was supported by CONACyT and SNI (Mexico). 


\begin{thebibliography}{99}


\bibitem{EWSBrev} S. Dawson et al., {\it The Higgs Hunter's Guide},
2nd ed., Frontiers in Physics Vol. {\bf 80} (Addison-Wesley,
Reading MA, 1990).


\bibitem{COSMOrev}
  L.~Bergstrom and A.~Goobar,
``Cosmology and particle astrophysics,''
{\it  Berlin, Germany: Springer (2004) 364 pp.}


\bibitem{hixlhc}
  J.~Ellis,
  ``Beyond the Standard Model at the LHC and Beyond,''
  arXiv:0710.0777 [hep-ph].



\bibitem{DMrev}
  G.~Bertone, D.~Hooper and J.~Silk,
  Phys.\ Rept.\  {\bf 405}, 279 (2005)
  [arXiv:hep-ph/0404175].


\bibitem{mssmrev}
  J.~R.~Ellis, S.~Heinemeyer, K.~A.~Olive, A.~M.~Weber and G.~Weiglein,
  JHEP {\bf 0708}, 083 (2007)
  [arXiv:0706.0652 [hep-ph]].


\bibitem{mssmXDM}
  J.~R.~Ellis, K.~A.~Olive, Y.~Santoso and V.~C.~Spanos,
  Phys.\ Rev.\  D {\bf 71}, 095007 (2005)
  [arXiv:hep-ph/0502001].

\bibitem{ewsbrev}
  C.~Csaki, J.~Hubisz and P.~Meade,
  arXiv:hep-ph/0510275.

\bibitem{exdimrev}
  M.~Quiros,
  arXiv:hep-ph/0606153;
  A.~Perez-Lorenzana,
  J.\ Phys.\ Conf.\ Ser.\  {\bf 18}, 224 (2005)
  [arXiv:hep-ph/0503177].
 

\bibitem{LTP}
  H.~C.~Cheng and I.~Low,
  JHEP {\bf 0408}, 061 (2004)
  [arXiv:hep-ph/0405243].


\bibitem{LKP}
  K.~Agashe and G.~Servant,
  Phys.\ Rev.\ Lett.\  {\bf 93}, 231805 (2004)
  [arXiv:hep-ph/0403143];

  G.~Servant and T.~M.~P.~Tait,
  Nucl.\ Phys.\  B {\bf 650}, 391 (2003)
  [arXiv:hep-ph/0206071].


\bibitem{HoloHiggs}
  R.~Contino, Y.~Nomura and A.~Pomarol,
  Nucl.\ Phys.\  B {\bf 671}, 148 (2003)
  [arXiv:hep-ph/0306259].



\bibitem{EffholoHIX}
  G.~F.~Giudice, C.~Grojean, A.~Pomarol and R.~Rattazzi,
  JHEP {\bf 0706}, 045 (2007)
  [arXiv:hep-ph/0703164].


\bibitem{EffholoKK} 
R.~Contino, T.~Kramer, M.~Son and R.~Sundrum,
  JHEP {\bf 0705}, 074 (2007)
  [arXiv:hep-ph/0612180].


 
\bibitem{holohixM}
  K.~Agashe, R.~Contino and A.~Pomarol,
  Nucl.\ Phys.\  B {\bf 719}, 165 (2005)
  [arXiv:hep-ph/0412089].

\bibitem{ourHMetal}
  J.~L.~Diaz-Cruz, J.~M.~Hernandez and J.~J.~Toscano,
  Mod.\ Phys.\ Lett.\  A {\bf 15}, 1377 (2000)
  [arXiv:hep-ph/9905335].

\bibitem{lightCUST}
  R.~Contino, L.~Da Rold and A.~Pomarol,
  Phys.\ Rev.\  D {\bf 75}, 055014 (2007)
  [arXiv:hep-ph/0612048].

\bibitem{ournextDM}J.L. Diaz-Cruz et al., work in progress.


\bibitem{wmapA}
  D.~N.~Spergel {\it et al.}  [WMAP Collaboration],
  Astrophys.\ J.\ Suppl.\  {\bf 170}, 377 (2007)
  [arXiv:astro-ph/0603449].

\bibitem{CDMexp}
  D.~S.~Akerib {\it et al.}  [CDMS Collaboration],
  Phys.\ Rev.\ Lett.\  {\bf 96}, 011302 (2006)
  [arXiv:astro-ph/0509259].

\bibitem{MinDM}
  M.~Cirelli, N.~Fornengo and A.~Strumia,
  Nucl.\ Phys.\  B {\bf 753}, 178 (2006)
  [arXiv:hep-ph/0512090].


\bibitem{miniHAWK} An example of such searches is the miniHAWk
project.


\end{thebibliography}
\end{document}